\documentclass{elsart5p}

\usepackage{natbib}

\usepackage{graphicx}
\usepackage{color}

\usepackage{amssymb}
\usepackage{listings}
\begin{document}

\begin{frontmatter}



\title{Unexpected Formation Modes of the First Hard Binary in Core
 Collapse}


\author[UTsukuba]{Ataru Tanikawa\corauthref{cor1}},
\ead{tanikawa@ccs.tsukuba.ac.jp}
\author[Princeton]{Piet Hut}, \&
\author[UTokyoTech]{Junichiro Makino}

\address[UTsukuba]{Center for Computational Science, University of
 Tsukuba, 1--1--1, Tennodai, Tsukuba, Ibaraki 305--8577, Japan}
\address[Princeton]{Institute for Advanced Study, Princeton, NJ 08540, USA}
\address[UTokyoTech]{Interactive Research Center of Science, Graduate
 School of Science and Engineering Tokyo Institute of Technology,
 2--12--1 Ookayama, Meguro, Tokyo 152-8551, Japan}

\corauth[cor1]{Corresponding author.}

\begin{abstract}
The conventional wisdom for the formation of the first hard binary in
core collapse is that three-body interactions of single stars form
many soft binaries, most of which are quickly destroyed, but
eventually one of them survives.  We report on direct N-body
simulations to test these ideas, for the first time.  We find that
both assumptions are often incorrect: 1) quite a few three-body
interactions produce a hard binary from scratch; 2) and in many cases
there are more than three bodies directly and simultaneously involved
in the production of the first binary.  The main reason for the
discrepancies is that the core of a star cluster, at the first deep
collapse, contains typically only five or so stars.  Therefore, the
homogeneous background assumption, which still would be reasonable
for, say, 25 stars, utterly breaks down.  There have been some
speculations in this direction, but we demonstrate this result
here explicitly, for the first time.
\end{abstract}

\begin{keyword}
Stellar dynamics \sep Method: $N$-body simulation \sep globular
clusters: general

\end{keyword}

\end{frontmatter}

\section{Introduction}

Nowadays, it is recognized that the dynamical evolution of a star
cluster is mostly understood as follows \citep{Heggie03}. Two-body
relaxation drives core collapse. The core collapse leads to high
stellar density region at the cluster center. Owing to the high
stellar density, binaries are formed through three-body encounters
\citep{Aarseth71,Heggie75,Hut85,Goodman93}. These binaries play
important roles for post-collapse evolution, triggering re-expansion
of the core, as well as gravothermal oscillations \citep{Bettwieser83,
Makino96}.

However, it has not been confirmed whether the binaries are actually
formed through three-body encounters, although it has been well
investigated how they evolve subsequently, after
formation\citep{Spurzem96}. Indirect evidence has been gathered
indicating that more than three stars might be involved
\citep{McMillan88}, but no direct observations have been published of
the core conditions that lead to the formation of the first binary, in
a simulation in which a cluster of single stars undergoes core
collapse. In contrast, later hardening of binaries has been described
in considerable detail by \cite{McMillan90} (MHM90).

In this paper, we have followed the dynamical evolution of star
clusters by performing $N$-body simulations, in order to catch the
moment of binary formation. Upon spotting the first binary, we have
then analyzed in detail how this binary formed, with how many other
stars involved, and in what kind of step-by-step process.

The structure of this paper is as follows. In section
\ref{sec:method}, we describe our method for $N$-body simulations and
our strategies for catching the moment of binary formation and for
analyzing the event. In section \ref{sec:results}, we show in detail
how the binaries are formed. In section \ref{sec:summary}, we sum up.

\section{Method}

\label{sec:method}

We have performed $N$-body simulations of star cluster evolution, up
to and slightly past core-collapse. The method for the $N$-body
simulations is described in section \ref{sec:n-body}. The timescales
of binary formation are very much smaller than those of cluster evolution
as a whole.  This means that the need for high-resolution data around
binary formation cannot be simply extended to a full recording of the
whole cluster evolution history with the same level of detail.  We
have to be quite selective in deciding what to store and what to throw
out.  Our strategy of data compression is described in section
\ref{sec:strategy}.

\subsection{$N$-body simulation}
\label{sec:n-body}

We have followed the dynamical evolution of five star clusters with
equal-mass stars without primordial binaries.  Our first three
clusters have about a thousand stars, $N=1k$, another has $N=4k$, and
the last one has $N=16k$, where $1k = 2^{10} = 1024$.  Their initial
stellar distributions are given by Plummer's model. Stars in the three
clusters with $N=1k$ are generated with different random seeds, which
we label here as seeds 1, 2, and 3.

For the $N$-body simulations, we have used GORILLA, an $N$-body simulation
code for star clusters \citep{Tanikawa09}. This code uses a
fourth-order Hermite scheme with individual timesteps \citep{Makino92}
for the time integration. The internal orbits of isolated binaries are
approximated as Kepler motion.  For further details, see TF09.

The accuracy parameter $\eta$, and its value at startup, $\eta_{\rm s}$,
were set to $0.01$ and $0.0025$, respectively. We have adopted the
standard $N$-body units, $G=M=r_{\rm v}=1$, \citep{Heggie86}, where
$G$ is the gravitational constant, and $M$ and $r_{\rm v}$
are respectively the total mass and virial radius of the cluster at
the start of the simulation.

We have performed the above $N$-body simulations on one node of the
FIRST cluster at the Center for Computational Sciences in University
of Tsukuba \citep{Umemura08}. Each node of the FIRST cluster is
equipped with Blade-GRAPE, a special purpose computer for $N$-body
simulations \citep{Sugimoto90,Makino03,Makino08}.

\subsection{Strategy to catch the moment of binary formation}
\label{sec:strategy}

\begin{figure*}
 \begin{center}
   \includegraphics[scale=1.5]{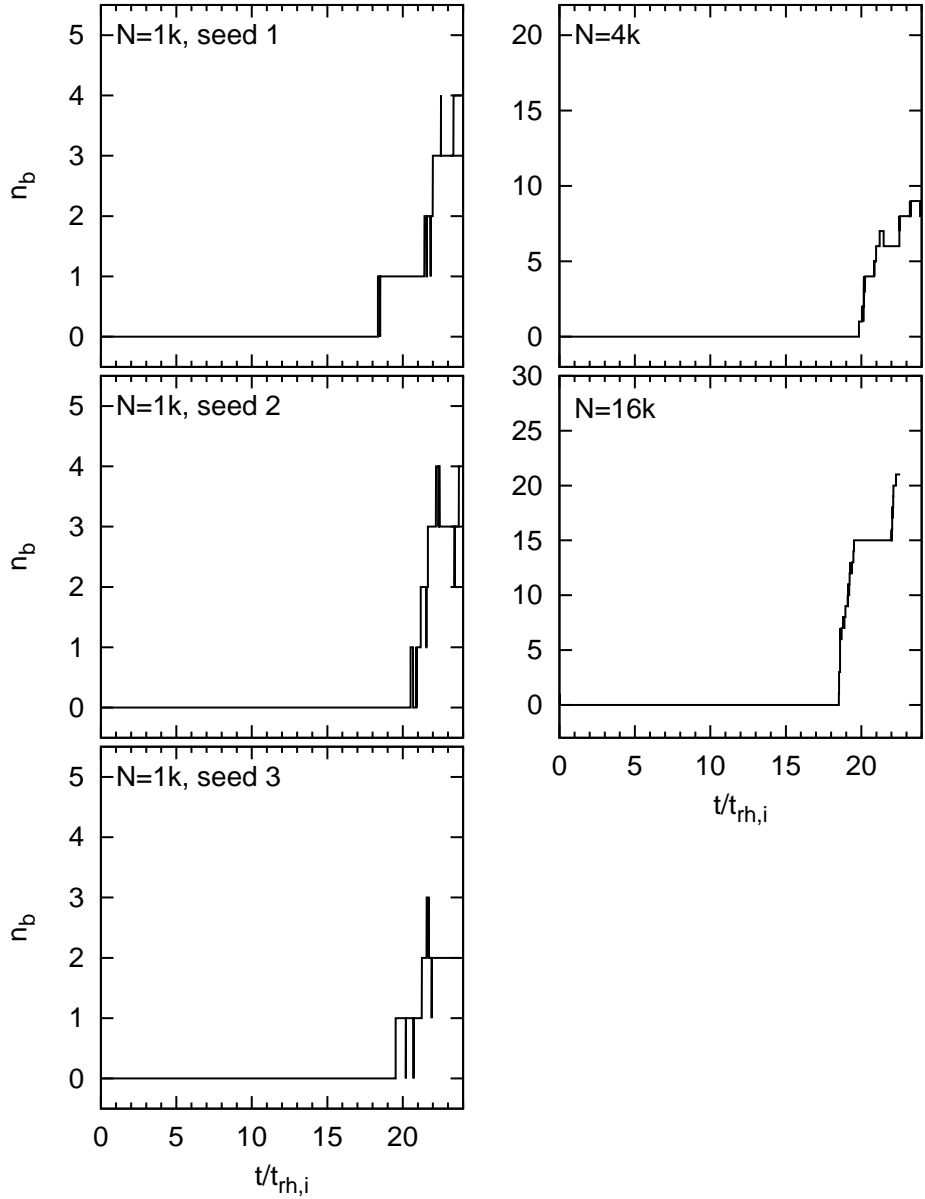}
 \end{center}
 \caption{Time evolution of the number of binaries with more than $10
   kT$ ($n_{\rm b}$). Time is scaled by initial half-mass relaxation
   time, defined in equation (\ref{eq:trh}).}
 \label{fig:nb}
\end{figure*}

We have caught the moment of binary formation manually, for each
simulation.  After determining the overall evolution of each cluster
by means of an $N$-body simulation as described in the previous
section, we have plotted the time evolution of the number of binaries,
shown in figure~\ref{fig:nb}.  To be specific, we have disregarded
binaries with only modest binding energy, since many of them dissolved
again before getting a chance to harden indefinitely.  We have chosen
our cut-off at $10 kT$, where $3/2kT$ is the average stellar kinetic
energy in the cluster, i.e. $1kT=1/(6N)$, at time $t=0$, at the start
of the simulation. The binding energy is expressed as
\begin{equation}
 e_{\rm b} = - \left( \frac{1}{2} \frac{m_1m_2}{m_1+m_2} v_{12}^2 -
 \frac{Gm_1m_2}{r_{12}} \right),
\end{equation}
where $m_1$ and $m_2$ are the masses of the binary components, and
$r_{12}$ and $v_{12}$ are the separation and relative velocity between
them. The time in these figures is scaled by the initial half-mass
relaxation time. The half-mass relaxation time is defined as
\begin{equation}
 t_{\rm rh} = 0.138 \frac{Nr_{\rm
     h}^{3/2}}{G^{1/2}M^{1/2}\log(0.4N)}, \label{eq:trh}
\end{equation}
where $r_{\rm h}$ is a half-mass radius of a cluster, and $r_{\rm h}
\simeq 0.8$ in the units at $t=0$.

After thus establishing the exact time of the formation of the first
$10 kT$ binary, we have done a complete rerun of the same simulation,
taking snapshots in much higher time resolution just before and after
the moment of binary formation, as illustrated in figure~\ref{fig:nb}.
Since the binary is formed in or around the cluster core, it is
sufficient to take snapshots only of the core regions.  The combination
of both temporal and spatial data reduction resulted in orders of
magnitude savings in memory needed to store the history of stellar
motions around the binary formation events.

We have taken snapshots at very small intervals of $0.01 t_{\rm cr,c}$,
only 1\% of the instantaneous core crossing time $t_{\rm cr,c}$, given by
\begin{equation}
 t_{\rm cr,c} = \frac{r_{\rm c}}{v_{\rm c}},
\end{equation}
where $r_{\rm c}$ is the core radius, and $v_{\rm c}$ is the stellar
velocity dispersion in the core.  For relatively large numbers of core
stars, we have only included those in our snapshots, but whenever the
number of stars in the core fell below a critical value $N_{\rm c,crit}$,
we have included the $N_{\rm c,crit}$ nearest stars around the density
center.  Our choice for $N_{\rm c,crit}$ depends on the number of
stars in the simulation:
$N_{\rm c,crit}=40$ ($N=1k$), $100$ ($N=4k$), and $1000$ ($N=16k$).
The density center is defined by the procedure described in
\cite{Casertano85}
(CH85), as modified by MHM90, in which a density around a given star is
defined by using the distance to the sixth nearest star.

For typical snapshots, most stars are not synchronized to the time of
that snapshot.  In those cases, we have derived their positions and
velocities through extrapolation, using a predictor given by Taylor
series up to the first-order time derivatives of their accelerations,
so-called jerk.

We summarize the data of the snapshot files, one for each rerun,
in table \ref{tab:snapshot}.  It is clear that the file size is
quite modest; the largest value is about $1$ Gbyte, for $N=16k$,
while smaller by almost two orders of magnitude for the other runs.
Reasons for the rapid increase of file size with $N$ are the fact
that core crossing time decreases for the same selection of time
interval in $N$-body time units, and the increase of the number of
stars that we save for larger $N$ values.  The fact that the $1k$ runs
have a file size that is not much smaller than the $4k$ run stems from
the more conservative approach with which we started the former runs,
saving data for a longer time span.

\begin{table}
 \caption{Summary for the snapshot file in the final reruns, and
   $t_{\rm rh,i}$ designates the initial half-mass relaxation time.}
 \label{tab:snapshot}
 \begin{center}
   \begin{tabular}{lc|rrr}
     $N$    & Seed & Time [$t_{\rm rh,i}$] & Snapshots & File size [byte] \\
     \hline
     $N=1k$ & $1$  & $18.323$ - $18.442$   & $14410$   & $19$  mega \\
     $N=1k$ & $2$  & $20.405$ - $20.524$   & $10172$   & $26$  mega \\
     $N=1k$ & $3$  & $19.490$ - $19.542$   & $3087$    & $7.7$ mega \\
     $N=4k$ & -    & $19.8049$ - $19.8072$ & $2142$    & $14$  mega \\
     $N=16k$ & -   & $18.5198$ - $18.5204$ & $19307$   & $1.2$ giga \\
   \end{tabular}
 \end{center}
\end{table}

\section{Results}
\label{sec:results}

\subsection{Binary formation and core evolution}

As is well known, binary formation is closely related to the evolution
of a cluster core. When core collapse proceeds, more than two stars
interact simultaneously because of the high stellar density in the
core.  As a result of these interactions, a binary is formed. In this
section, we investigate the details of core evolution around the time
of binary formation.

\begin{figure*}
 \begin{center}
   \includegraphics[scale=1.2]{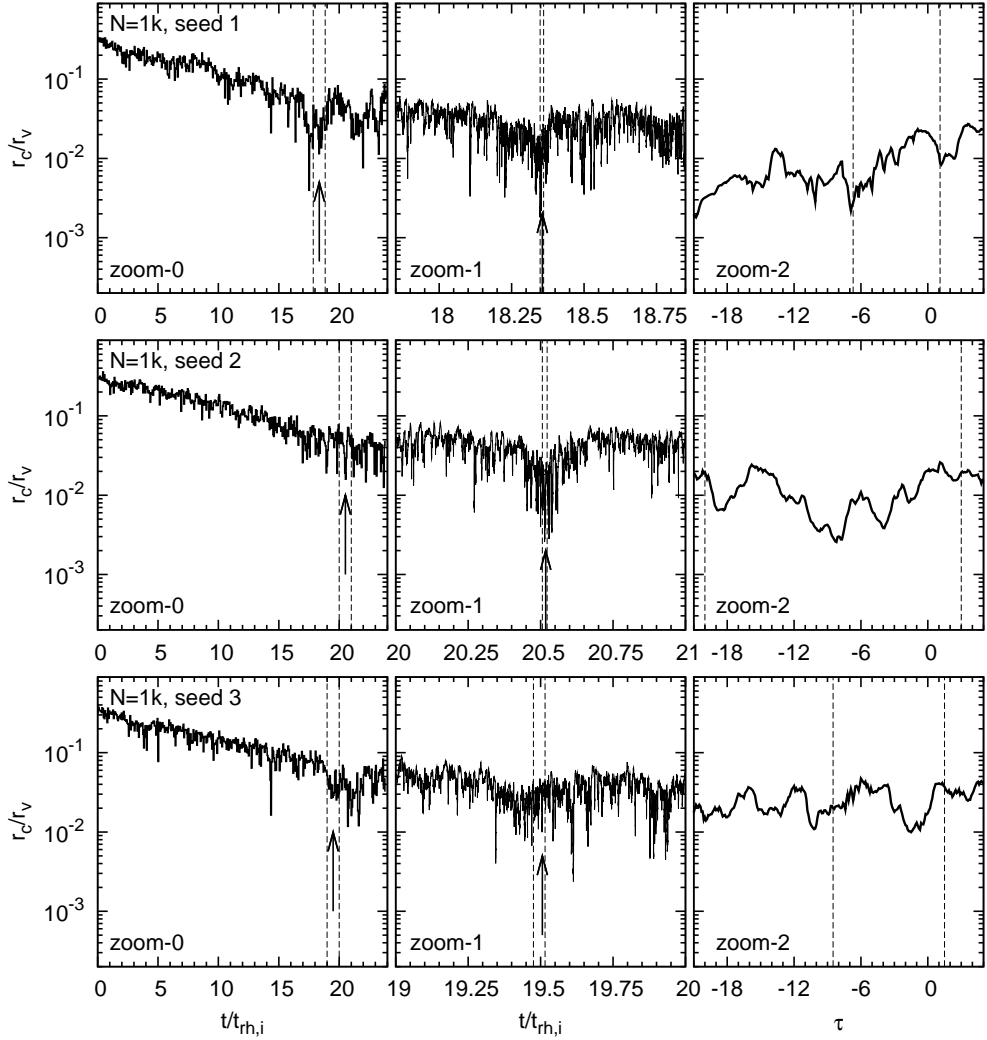}
 \end{center}
 \caption{Time evolution of core radii in the $N=1k$ runs with seed
   $1$, $2$, and $3$. The core radii in zoom-$0$ panels are calculated
   at each $1$ $N$-body time unit ($5.9 \times 10^{-2} t_{\rm
     rh}$). Those in the other panels are calculated at each $0.1
   t_{\rm cr,c}$. The width of the time in panels zoom-$1$ is $1
   t_{\rm rh,i}$. The range of the time in panel zoom-$n$ is the same
   as that between two vertical dashed lines in panel
   zoom-($n-1$). The ranges between two vertical dashed lines in
   panels zoom-2 are the same as those of the time in
   figure~\ref{fig:sep_n01k_seed01}, \ref{fig:sep_n01k_seed02}, and
   \ref{fig:sep_n01k_seed03}. The arrows and the time $\tau = 0$ show
   the time when the first binary is formed in each run.}
 \label{fig:rc_n01k}
\end{figure*}

\begin{figure*}
 \begin{center}
   \includegraphics[scale=1.2]{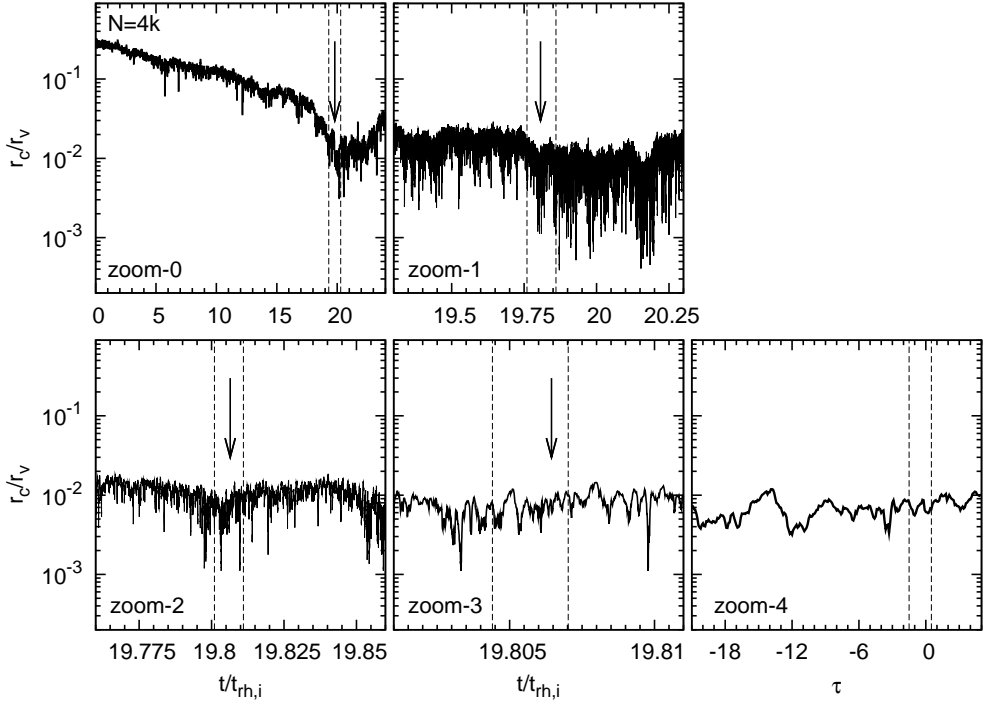}
 \end{center}
 \caption{Time evolution of core radii in the $N=4k$ run. The core
   radius in zoom-$0$ panel is calculated at each $1$ $N$-body time
   unit ($1.8 \times 10^{-2} t_{\rm rh}$). Those in the other panels
   are calculated at each $0.1 t_{\rm cr,c}$. The width of the time in
   panels zoom-$1$ is $1 t_{\rm rh,i}$. From zoom-$1$ to zoom-$3$, the
   time is zoomed out by ten times step by step. The meanings of the
   dashed lines are the same as those in figure~\ref{fig:rc_n01k}. The
   range between the two vertical dashed lines in panel zoom-$4$ is
   the same as that of the time in
   figure~\ref{fig:sep_n04k_seed01}. The arrows and the time $\tau =
   0$ show the time when the first binary is formed.}
 \label{fig:rc_n04k}
\end{figure*}

\begin{figure*}
 \begin{center}
   \includegraphics[scale=1.2]{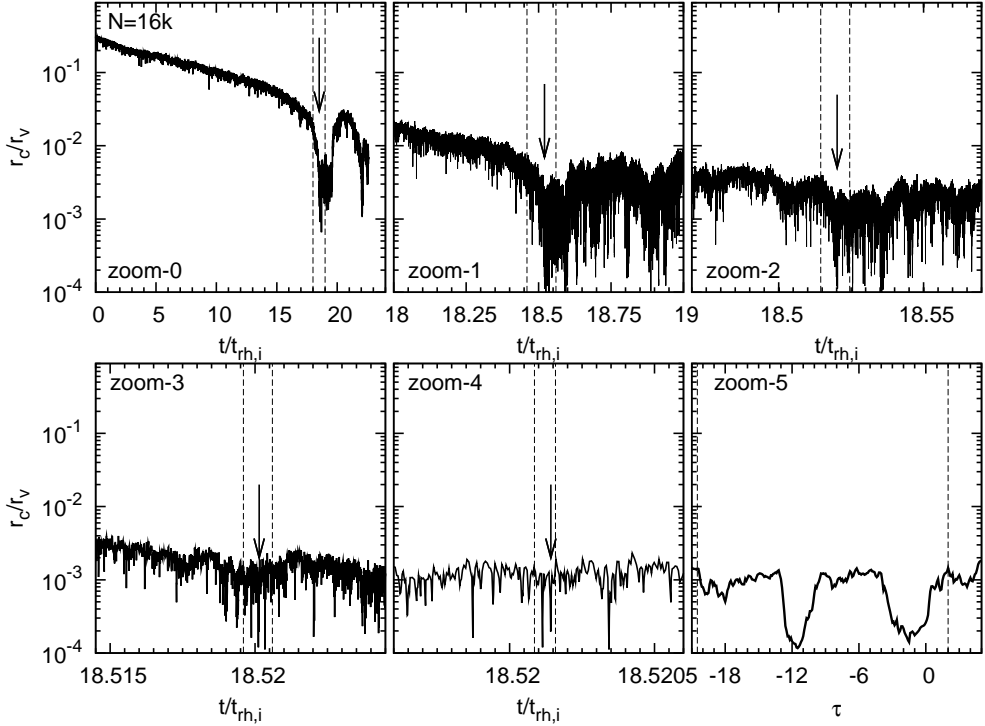}
 \end{center}
 \caption{Time evolution of core radii in the $N=16k$ run. The arrows
   and the time $\tau = 0$ show the time when the first binary is
   formed. The core radius in zoom-$0$ panel is calculated at each $1$
   $N$-body time unit ($5.4 \times 10^{-3} t_{\rm rh}$). Those in
   panels zoom-$1$, $2$, $3$, and $4$ are calculated at each $1 t_{\rm
     cr,c}$, and those in panel zoom-$5$ at each $0.1t_{\rm
     cr,c}$. The width of the time in panels zoom-$1$ is $1 t_{\rm
     rh,i}$. From zoom-$1$ to zoom-$4$, the time is zoomed out by ten
   times step by step. The meanings of the dashed lines are the same
   as those in figure~\ref{fig:rc_n01k}. The range between the two
   vertical dashed lines in panel zoom-$4$ is the same as that of the
   time in figure~\ref{fig:sep_n16k_seed01}.}
 \label{fig:rc_n16k}
\end{figure*}

Figure~\ref{fig:rc_n01k} -- \ref{fig:rc_n16k} show the time evolution
of core radii of the clusters, and the moment of first binary
formation. The core radii are derived in the same way as CH85
modified by MHM90, in which a density around a given star is defined
from the sixth nearest star. In the panels with maximum zoom factor
(zoom-$2$ for $N=1k$; zoom-$4$ for $N=4k$; and zoom-$5$ for $N=16k$)
time is scaled by the instantaneous value of the core crossing time.
We can express this `co-moving time' $\tau$ as
\begin{equation}
 \tau = \int \frac{dt}{t_{\rm cr,c}}.
\end{equation}
$\tau$ effectively presents the accumulation of time on a clock with
variable speed, in such a way that each increment in time is scaled by
the value of the core crossing time at that particular moment. In this
way, the very brief periods of deepest core collapse are guaranteed to
be magnified so as to be completely resolved in time.  The first
binaries are formed during the time interval between the two dashed
lines in the panels with maximum zoom factor.

As expected, the time at which the first binaries are formed is
comparable for the different runs, when expressed in $N$-body time
units: the first binary in each run is formed between $18 t_{\rm
  rh,i}$ and $21 t_{\rm rh,i}$. At the time of the formation of the
first $10 kT$ binary, the core radii are $2 \times 10^{-3}$ -- $2
\times 10^{-2}$ for the $N=1k$ runs with seed 1 and 2, $1 \times
10^{-2}$ -- $4 \times 10^{-2}$ for the $N=1k$ run with seed 3, $5
\times 10^{-3}$ -- $1 \times 10^{-2}$ for the $N=4k$ run, and $1
\times 10^{-4}$ -- $1 \times 10^{-3}$ for the $N=16k$ run (see the
panels with maximum zoom factor in figure~\ref{fig:rc_n01k} --
\ref{fig:rc_n16k}).

When inspecting the zoom-$1$ panels of figures~\ref{fig:rc_n01k},
\ref{fig:rc_n04k}, and \ref{fig:rc_n16k}, the core radii around the
time pointed at by the arrows take on locally minimal values during
intervals of $0.1 t_{\rm rh,i}$ before and after these times. This can
be further investigated by following the time evolution of the maximum
values of the core radii during fluctuations on smaller timescales
than the initial half-mass relaxation time. The maximum values for the
core radii, around the times pointed at by the arrows, are $2 \times
10^{-2}$ for the $1k$ runs, $1 \times 10^{-2}$ for the $4k$ run, and
$2 \times 10^{-3}$ for the $16k$ run. In contrast, corresponding
values at other times during this $0.1 t_{\rm rh,i}$ period are $3
\times 10^{-2}$ for the $1k$ runs, $2 \times 10^{-2}$ for the $4k$
run, and $4 \times 10^{-3}$ for the $16k$ run.

Looking in further detail at the core radii around the time of binary
formation, there is no precise correlation between binary formation
and core radii.  In fact, core radii at times during which no binary
is formed are just about as small as core radii at binary formation
time, something that is especially clear in the $4k$ and $16k$ runs
(see zoom-$3$ and zoom-$4$ of figure~\ref{fig:rc_n04k}, and zoom-$4$
of figure~\ref{fig:rc_n16k}, respectively). Core radii are almost
constant during $1 \times 10^{-2} t_{\rm rh,i}$ in the case of the
$4k$ run (see zoom-$3$ of figure~\ref{fig:rc_n04k}) and $1 \times
10^{-3} t_{\rm rh,i}$ in the case of the $16k$ run (see zoom-$4$ of
figure~\ref{fig:rc_n16k}). On the other hand, the binaries are formed
during intervals much less than $2.5 \times 10^{-3}t_{\rm rh,i}$
($4k$) and $5 \times 10^{-5} t_{\rm rh,i}$ ($16k$).  The timescale on
which core radii small enough to form a binary (about $1\times
10^{-2}$ for $4k$ and $1 \times 10^{-3}$ for $16k$) persist is several
times longer than binary formation timescale.  We are clearly dealing
with events with low probability, even at these high densities,
occurring stochastically.

On the way toward binary formation, the core radii sometimes become
smaller than any other time by a factor of ten, for example $\tau =
-7$ in the $N=1k$ run with seed 1, $\tau = -7$ in the $N=1k$ run with
seed 2, $\tau = -12$ and $-1$ in the $N=16k$ run. These decrease of
core radii is not correlated with binary formation. Such decrease of
core radii occurs at the other time.

\begin{figure*}
 \begin{center}
   \includegraphics[scale=1.5]{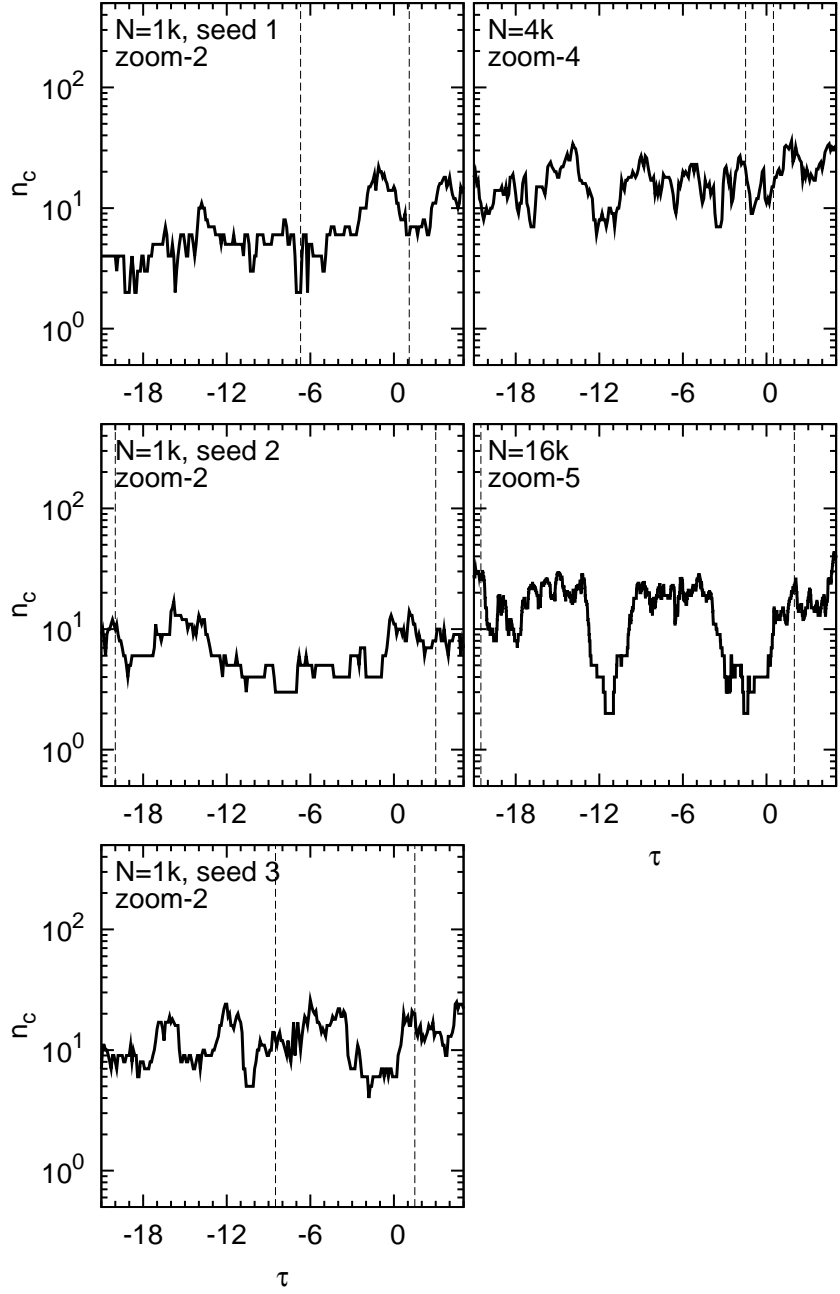}
 \end{center}
 \caption{Time evolution of the number of stars in core at around
   binary formation in the $N=1k$ runs with seed 1, 2, and 3, $N=4k$
   run, and $N=16k$ run.}
 \label{fig:nc}
\end{figure*}

Figure~\ref{fig:nc} shows the time evolution of the number of stars in
the core around the time of binary formation. Note that the numbers of
stars in the core around that time fluctuates by roughly a factor or
$10$, with significant differences between different runs. This in
itself already indicates that a homogeneous background density
assumption for three-body binary formation breaks down.  As we will
now see in detail, binaries are indeed formed by processes other than
simple three-body encounters.

\subsection{Hard Binary Formation Histories}
\label{sec:history}

For more than 35 years, the conventional wisdom for the formation of
the first hard binary in core collapse has postulated a relatively
straightforward picture.  While the density of the core increases,
near-simultaneous three-body encounters become more and more likely.
These can lead to the formation of binaries, where the third star
picks up the excess energy, leaving the other two stars bound.
Subsequent encounters are most likely to destroy these newly formed
soft binaries, but occasionally a soft binary hardens against the
odds, and then becomes stable when its binding energy exceeds several
$kT$ \citep{Heggie75,Hut83,Goodman93}.

We are now in a position to check this conventional wisdom, using the
information we have gathered about orbits of stars and the time
evolution of the pair-wise distances between stars in the core.

\begin{figure*}
 \begin{center}
   \includegraphics[scale=1.5]{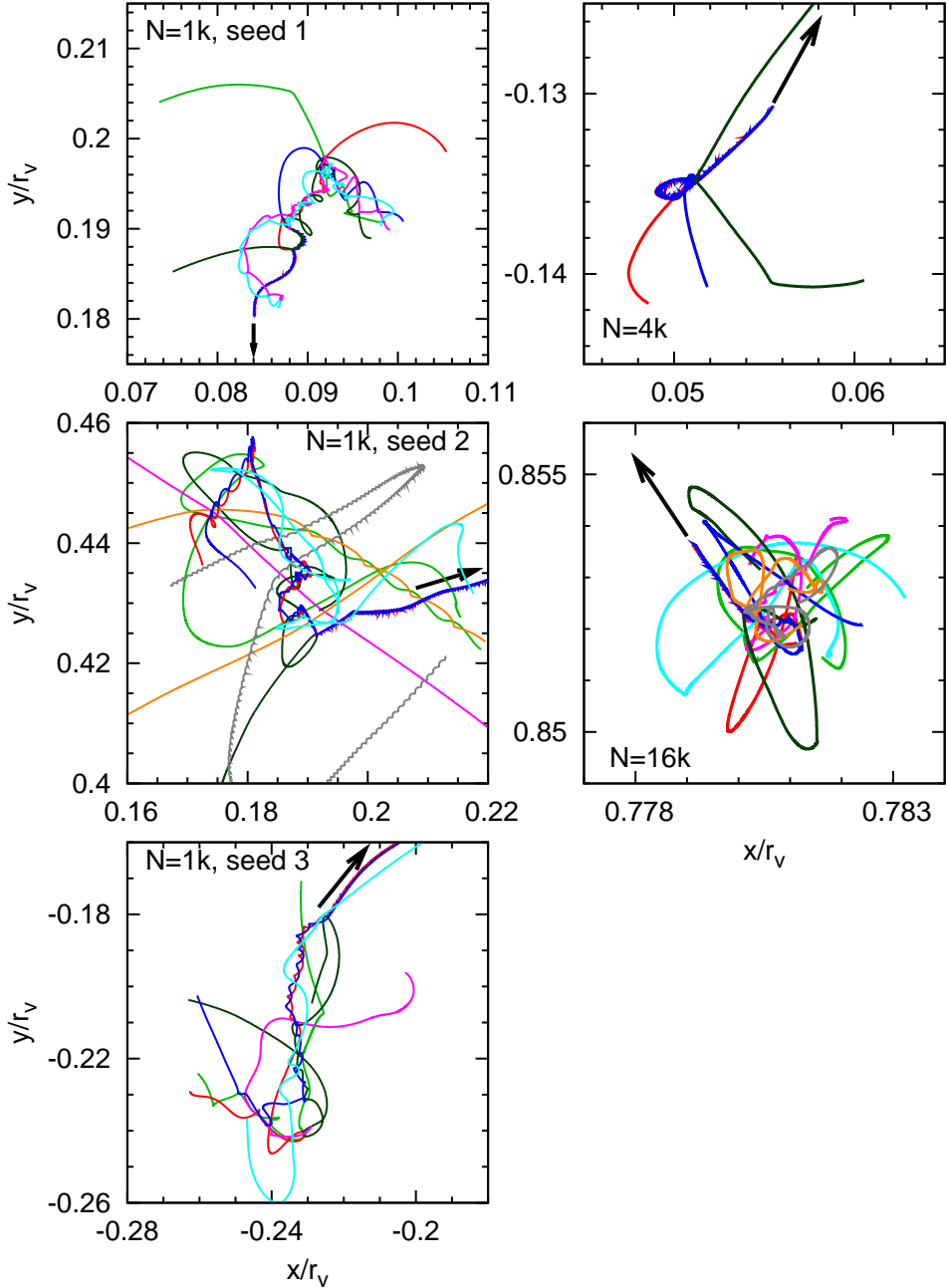}
 \end{center}
 \caption{ Orbits of stars involving the binary formation in the
   $N=1k$ run with seed 1. The stars whose orbits are drawn by red and
   blue curves finally become the binary components, and their
   directions are indicated by black arrows. The time ranges are the
   same as those in panels zoom-$2$, $4$, and $5$ in
   figure~\ref{fig:rc_n01k}, \ref{fig:rc_n04k}, and \ref{fig:rc_n16k},
   respectively.}
 \label{fig:xyz}
\end{figure*}

First, let us have a look at the orbits of stars involving binary
formation in figure~\ref{fig:xyz}. Apparently, the binary in the
$N=4k$ run, which presents an unusually simple formation case, is
formed through an encounter among three single stars. The red, blue,
and dark green stars simultaneously approach each other. Subsequently,
the red and blue stars compose of a binary, and the dark green star is
ejected from them. This indeed corresponds to the classical picture
for binary formation. On the other hand, many stars involve the
formation of the binaries in all the other runs. Their orbits are too
complicated to determine the mechanism of binary formation from these
figures, even though they already suggest a mechanism quite different
from the classical picture.

\begin{figure*}
 \begin{center}
   \includegraphics[scale=1.35]{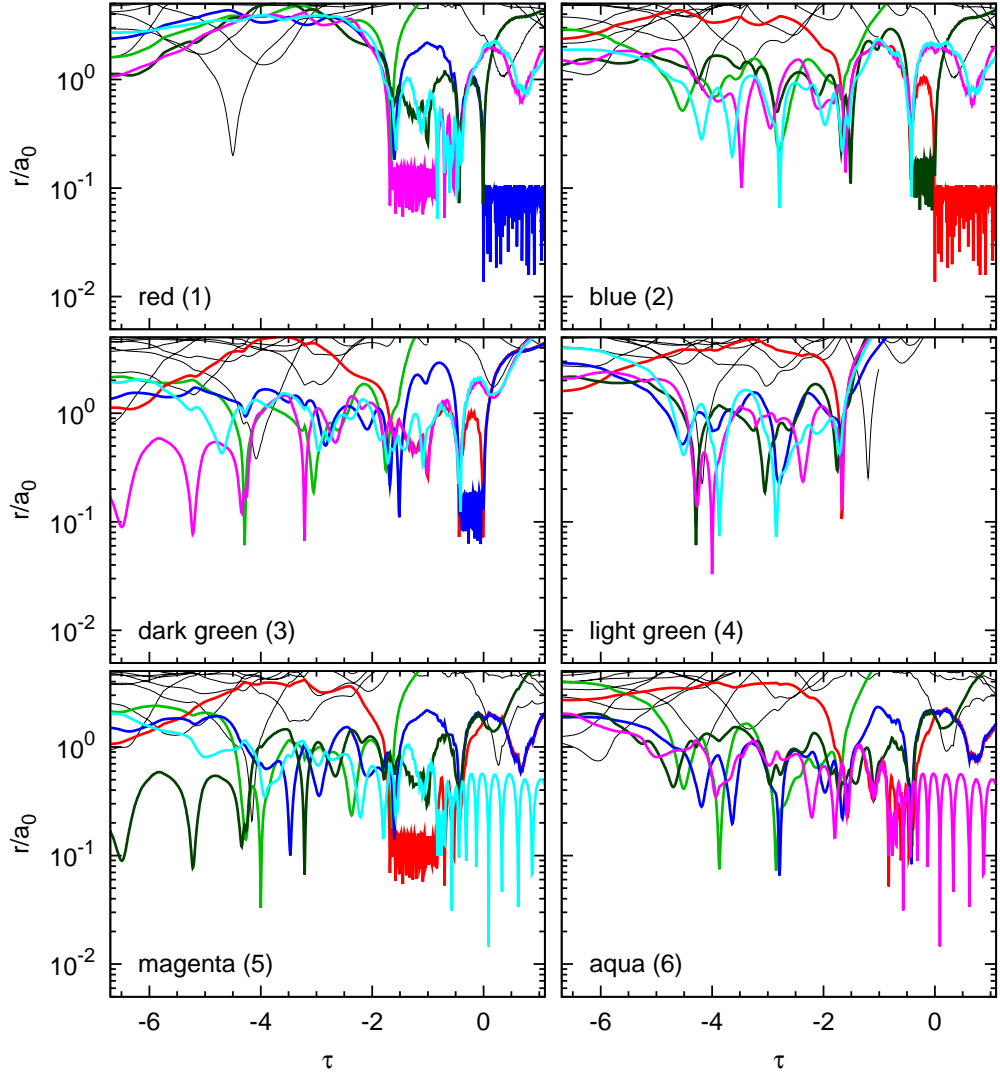}
 \end{center}
 \caption{Time evolution of separations among stars involving binary
   formation in the $N=1k$ run with seed 1. The colors are the same as
   those in figure~\ref{fig:xyz}. The black curves indicate stars
   which do not involve the binary formation. For each star, following
   the color, the number between parentheses refers to the numbers
   used in the discussion in section \ref{sec:history}.}
 \label{fig:sep_n01k_seed01}
\end{figure*}

\begin{figure*}
 \begin{center}
   \includegraphics[scale=1.35]{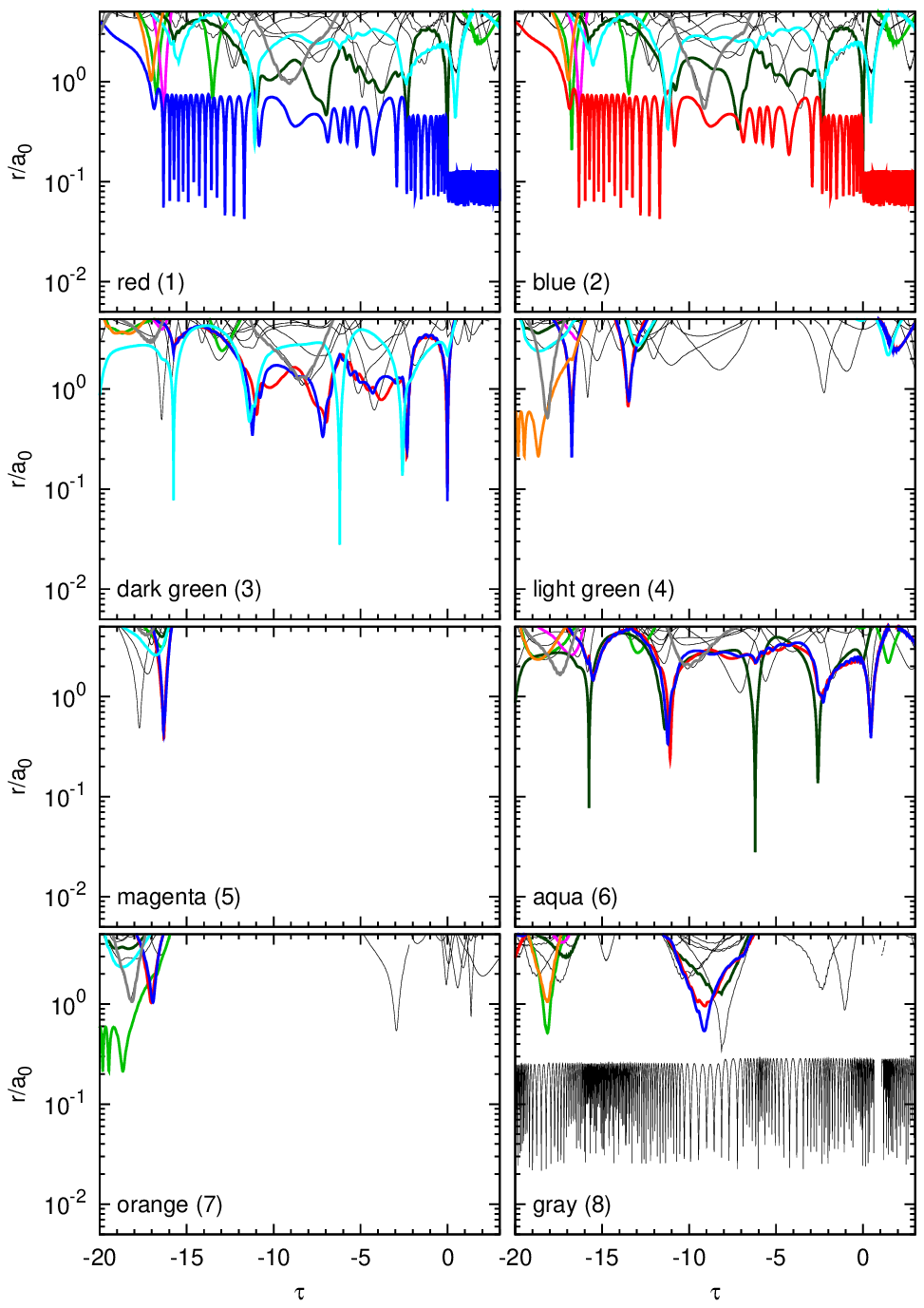}
 \end{center}
 \caption{Time evolution of separations among stars involving binary
   formation in the $N=1k$ run with seed 2.}
 \label{fig:sep_n01k_seed02}
\end{figure*}

\begin{figure*}
 \begin{center}
   \includegraphics[scale=1.35]{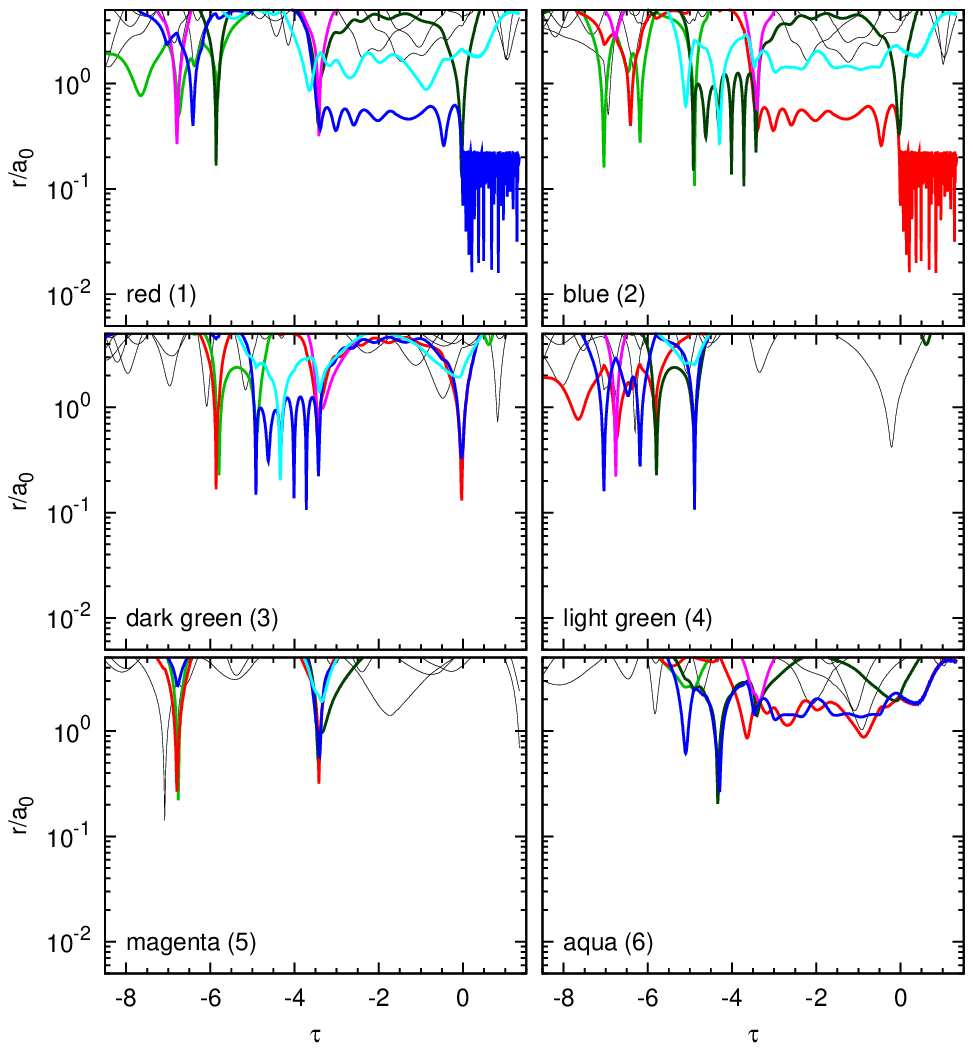}
 \end{center}
 \caption{Time evolution of separations among stars involving binary
   formation in the $N=1k$ run with seed 3.}
 \label{fig:sep_n01k_seed03}
\end{figure*}

\begin{figure*}
 \begin{center}
   \includegraphics[scale=1.5]{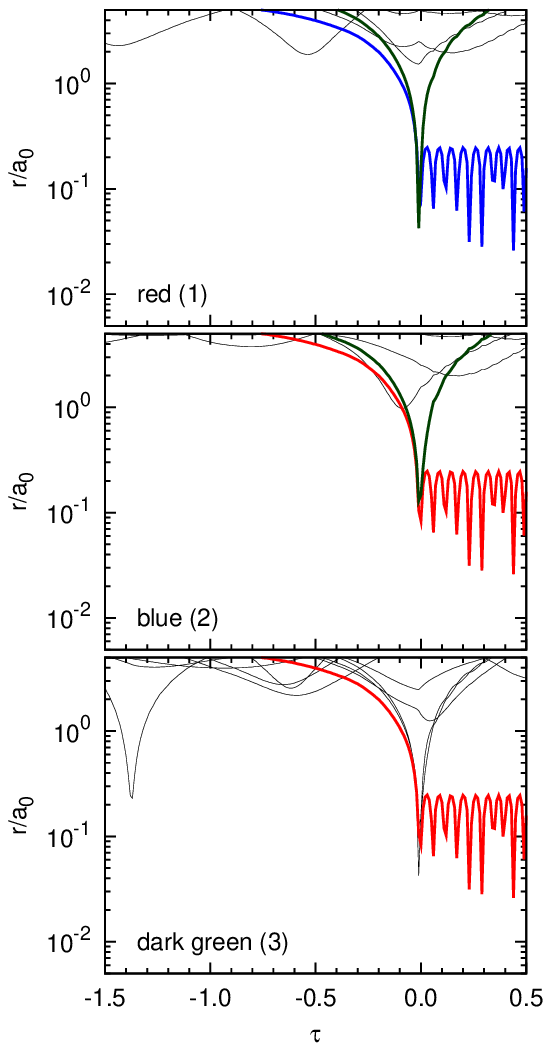}
 \end{center}
 \caption{Time evolution of separations among stars involving binary
   formation in the $N=4k$ run.}
 \label{fig:sep_n04k_seed01}
\end{figure*}

\begin{figure*}
 \begin{center}
   \includegraphics[scale=1.35]{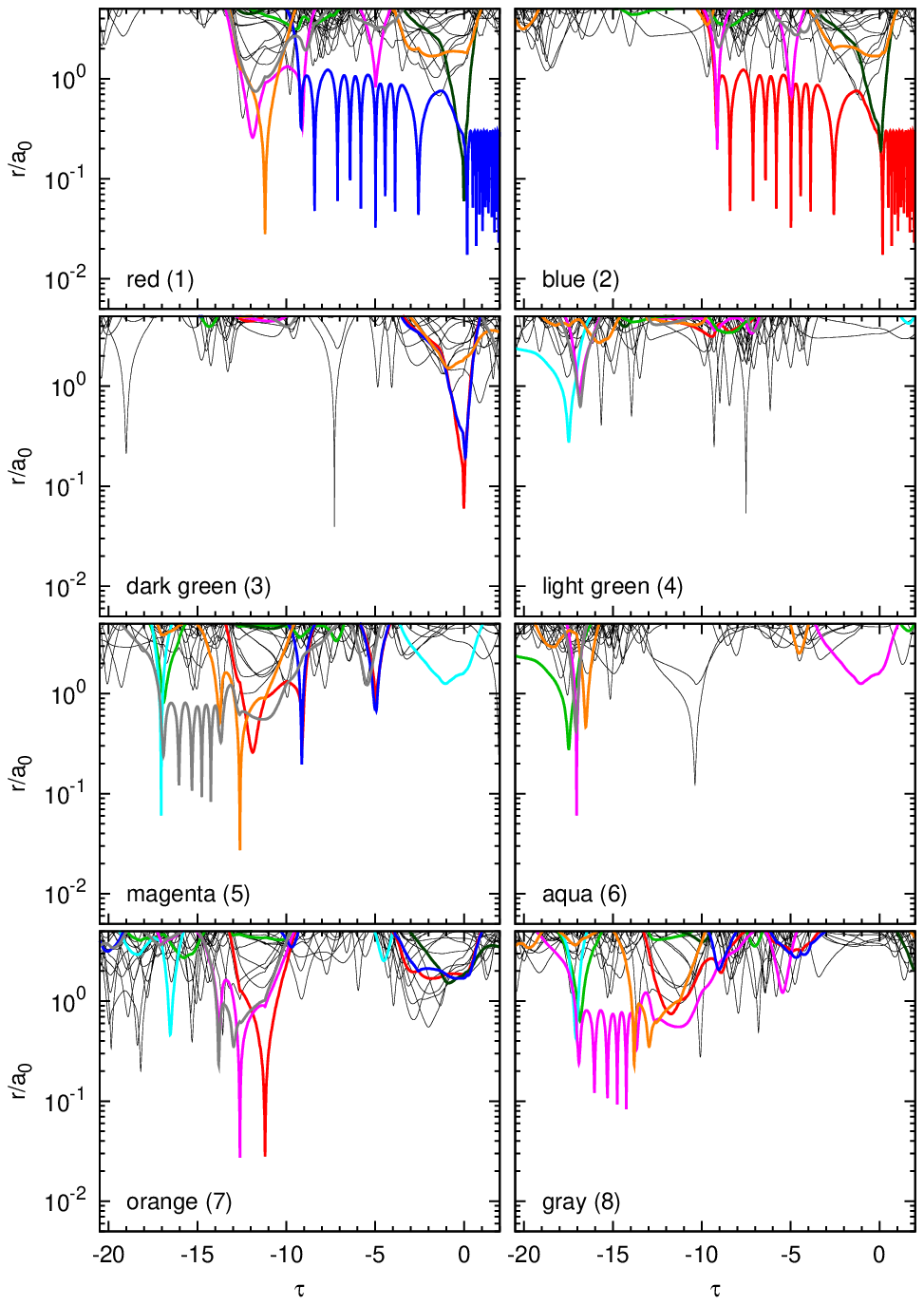}
 \end{center}
 \caption{Time evolution of separations among stars involving binary
   formation in the $N=16k$ run.}
 \label{fig:sep_n16k_seed01}
\end{figure*}

In order to find out more about what is actually happening, we have
plotted the time evolution of the pair-wise distances between all
stars in the core, during a very small period of time just before
binary formation. Figures~\ref{fig:sep_n01k_seed01} --
\ref{fig:sep_n16k_seed01} show the interactions of stars shown in
figure~\ref{fig:xyz} in further detail. Each separate panel shows the
distances of one particular star to all other stars in the core.  For
each curve in a given panel, the color of the curve indicates the
identity of the star to which the pair-wise distance is plotted.  When
some of the curves are intermittent, for example the nearest black
curve in the gray panel of figure~\ref{fig:sep_n01k_seed02}, it is an
indication that one or another of the two stars in a pair have
temporarily moved outside the cluster core. As described in section
\ref{sec:strategy}, we have not recorded the data for stars outside
the cluster core.

We now describe the evolution histories in detail. We will start with
figure~\ref{fig:sep_n01k_seed01}, where the results for the $N=1k$ run
are given, for seed 1.  Just before $\tau = -1.5$, an encounter
involving four stars (1, 2, 4, 5) leads directly to the formation of a
hard binary (1, 5).  A little later, just after $\tau = -1$, that
binary dissolves in a complicated dance involving five stars (1, 2, 3,
5, 6), and soon thereafter, around $\tau = -0.5$, a new hard binary
(2,3) emerges from this dance.  Finally, at $\tau = 0$, a direct
three-body exchange leads to the first hard binary (1, 2) with an
energy $\lg 10kT$.

Figure~\ref{fig:sep_n01k_seed02} shows the results for the $N=1k$ run
with seed 2.  At the start of the interval depicted, there already is
a hard binary (8) that was formed earlier; since this binary is only
slightly perturbed during this interval, retaining its identity, we
have not numbered its two members separately: number 8 indicates both
stars as a unit.  Also present at the start of the interval is a
mildly hard binary (4, 7) which by $\tau = -18$ dissolves during an
encounter with three other stars (1, 2, 5).  Out of this five-body
interaction emerges, soon thereafter around $\tau = -17$, another
mildly hard binary (1, 2).  This new binary undergoes various
perturbations by passing stars, hardening gradually at first.  Then,
at $\tau = 0$, a close encounter with a single star (3) leads to a
sudden increase in binding energy, by a factor of two to three.

Figure~\ref{fig:sep_n01k_seed03} shows a similar picture for the
$N=1k$ run with seed 3.  Around $\tau = -5$, a three-body encounter
(2, 3, 4) leads to the formation of a mildly hard binary (2, 3).  A
little later, at $\tau = -3.5$, a simultaneous encounter between this
binary and two single stars (1 and 5) leads to the formation of a
significantly harder binary (1, 2).  At $\tau = 0$, a close encounter
with a single star (3) leads to a sudden increase in binding energy,
by about a factor of two.

Figure~\ref{fig:sep_n04k_seed01} shows the $N=4k$ run. In contrast to
the preceding three cases, here hard binary formation is much simpler.
At $\tau = 0$ three stars (1, 2, 3) have a simultaneous close
encounter, leading directly to the formation of the first hard binary
(1, 2) with an energy $\lg 10kT$.

Figure~\ref{fig:sep_n16k_seed01}, for the $N=16k$ run, is again far
more complicated.  At $\tau = -17$, four single stars (4, 5, 6, 8)
meet, producing a mildly hard binary (5, 8).  At $\tau = -14$ the
binary dissolves in a many-body dance involving several other stars
that keep interacting until at $\tau = -9$ a new binary (1, 2) is
formed; at the last step in its formation mainly three stars (1, 2, 5)
were involved.  Finally, at $\tau = 0$, a close encounter with a
single star (3) leads to a sudden increase in binding energy, by a
factor of two to three.

We conclude from these five cases that direct three-body binary
formation is relatively rare.  In the one case that we observed, in
figure~\ref{fig:sep_n04k_seed01}, it led to a binary with a binding
energy of more than $10kT$, a very different picture than the gradual
hardening of a soft binary.

In the other four cases that we studied, the situation was quite
a bit more complicated.  In all of those cases, there were episodes
in which at least four or five stars were simultaneously involved
in close encounters that played a role in the formation of the
first really hard binary.

\section{Summary}
\label{sec:summary}

In this paper, we have shown that core collapse of star clusters gives
rise to the formation of hard binaries via far more complicated
processes than have been hitherto considered.  Most importantly, the
traditional approximation of binary formation through simultaneous
encounters of three bodies offers a vastly oversimplified picture of
what really happens.  Typically, more stars are strongly involved,
dynamically, in a complex multi-body dance that in no way can be
disentangled to lead to a hierarchy with only three stars in the center.

We have found only one exception, among the five cases studied in
great detail.  Even in that case, the traditional picture of slow
hardening of an initially quite soft binary doesn't apply.  Rather,
that three-body process of binary formation produced a rather hard
binary very quickly, without subsequent hardening encounters with other
stars.

We can point to a single reason for the fact that the two core
assumptions (3-body encounters, and gradual hardening) of the
traditional scenario for hard binary formation turn out to be
unrealistic.  The main reason for the break-down of those assumptions
is the very small core at the moment of core collapse.  What we have
found is that, typically, the core at that time contains less than 10
stars, and often only 5 or so stars -- so few that a precise measure
of the actual number of stars becomes difficult.  Given the discrete
nature of core stars, the use of continuum properties like density for
a population of only 5 stars is obviously rather problematic.

If core bounce would occur at somewhat larger core sizes, with cores
containing twenty to thirty stars or more, the assumptions of
three-body encounters and gradual hardening may well have been
correct.  Following the path breaking analytical investigation of the
role of binaries by \cite{Heggie75} in the seventies, subsequent
analytical work in the eighties seems to suggest that core sizes
during core bounce remained large enough to ensure Heggie's
assumptions.  For example, from table 2 of \cite{Goodman84},
equilibrium values of core sizes between $N=1k$ and $N=16k$ range
between twenty and sixty stars.  A more detailed analysis in
\cite{Goodman87} gave rise to his equation (IV.8) suggesting a core
size at deep bounce large enough for the core to contain dozens of
stars.

Our simulations, in contrast, suggest that deep core collapse causes
the number of stars in the core to dip below ten in all cases,
independent of the number $N$ of stars for the cluster as a whole.
During the final stages before the deepest collapse, the remainder
of the cluster is effectively frozen, on the local time scale of the
most central regions, and even the order of magnitude of $N$ may
become irrelevant.  We have demonstrated this for the range $N = 1k$
to $N = 16k$, and we predict a very similar behavior for larger
$N$ values, up to $N = 10^6$.

It is an interesting question, why the analytic estimates for the core
size at the time of core bounce have overestimated the number of stars
in the core $N_{\rm c,min}$ by almost an order of magnitude. Our
results indicate $N_{\rm c,min} \sim 5$ while the analytic estimates
have predicted results more like $N_{\rm c,min} \sim 30$. One possible
reason could be the breakdown of the instantaneous approximation for
interactions, used in analytic treatments. With the timescale for
formation and subsequent hardening of binaries being comparable to the
timescale of core contraction, we may need to extend the analytical
approach. We are currently looking into these questions in more
detail, continuing the investigations reported here.

\section*{Acknowledgment}

A. Tanikawa thanks Takashi Okamoto for helpful advice. Numerical
simulations have been performed with computational facilities at the
Center for Computational Sciences in University of Tsukuba. This work
was supported in part by the FIRST project based on the Grants-in-Aid
for Specially Promoted Research by MEXT (16002003), by Grant-in-Aid
for Scientific Research (S) by JSPS (20224002), and by
KAKENHI(21244020). Part of the work was done while two of the authors
(P. Hut and J. Makino) visited the Center for Planetary Science
(CPS). We are grateful for their hospitality.

\end{document}